\documentstyle[12pt]{article}
\begin{document}
\begin{titlepage}
\hspace*{\fill} ULB--TH--94/07\\
\hspace*{\fill} NIKHEF--H 94--15\\

\vspace{2.5cm}
\begin{centering}

{\huge Local BRST cohomology in the antifield formalism: II. Application to
Yang-Mills theory}

\vspace{2cm}
{\large Glenn Barnich$^{1,*}$, Friedemann Brandt $^{2,**}$ and \\
Marc Henneaux$^{1,***}$}\\
\vspace{1cm}
$^1$Facult\'e des Sciences, Universit\'e Libre de Bruxelles,\\
Campus Plaine C.P. 231, B-1050 Bruxelles, Belgium\\
$^2$NIKHEF-H, Postbus 41882, 1009 DB Amsterdam,\\ The
Netherlands\\

\end{centering}

\vspace{2.5cm}

{\footnotesize \hspace{-0.6cm}($^*$)Aspirant au Fonds National de la
Recherche
Scientifique (Belgium).\\
($^{**}$)Supported by Deutsche Forschungsgemeinschaft.\\
($^{***}$)Also at Centro de Estudios
Cient\'\i ficos de Santiago, Chile.}
\end{titlepage}

\begin{abstract}
Yang-Mills models with compact gauge group coupled to matter fields
are considered. The general tools developed in a companion paper are
applied to compute the local cohomology
of the BRST differential $s$ modulo the exterior spacetime derivative $d$
for all values of the ghost number, in the space of polynomials
in the fields, the ghosts, the
antifields (=sources for the BRST variations) and their derivatives.
New solutions to the consistency conditions $sa+db=0$ depending
non trivially on the antifields are exhibited. For a semi-simple gauge
group, however, these new solutions arise only at ghost number two or higher.
Thus at ghost number zero or one, the inclusion of the antifields does not
bring in new solutions to the consistency condition $sa+db=0$ besides the
already known ones.
The analysis does not use power counting and is purely
cohomological.  It can be easily extended
to more general actions containing higher derivatives of the
curvature, or Chern-Simons terms.
\end{abstract}

\pagebreak

\def\qed{\hbox{${\vcenter{\vbox{
   \hrule height 0.4pt\hbox{\vrule width 0.4pt height 6pt
   \kern5pt\vrule width 0.4pt}\hrule height 0.4pt}}}$}}
\newtheorem{theorem}{Theorem}[sectionc]
\newtheorem{lemma}{Lemma}
\newtheorem{definition}{Definition}
\newtheorem{corollary}{Corollary}
\newcommand{\proof}[1]{{\bf Proof.} #1~$\qed$.}
\renewcommand{\theequation}{\thesection.\arabic{equation}}
\renewcommand{\thetheorem}{\thesection.\arabic{theorem}}
\renewcommand{\thelemma}{\thesection.\arabic{lemma}}

\section{Introduction}
\setcounter{equation}{0}
In a previous paper \cite{BBH}, referred to as I, we have derived
general theorems on the local cohomology of the BRST differential $s$
for a generic gauge theory. We have discussed in particular how it is
related to the local cohomology of the Koszul-Tate differential $\delta$
and have demonstrated vanishing theorems for the cohomology
$H_k(\delta|d)$ under various conditions.
In the present paper, we apply the general results of I to
Yang-Mills models with compact gauge group
and provide the explicit list of all the
non-vanishing BRST groups $H^k(s|d)$ for those models.

It has been established on general grounds that the
groups $H^k(s)$ and $H^k(s|d)$ are respectively given by
\begin{eqnarray}
H^k(s)\simeq\cases{H^k\left(\gamma,H_0(\delta)\right)\quad k \geq 0
\label{1.1}\cr
0 \quad k<0\cr}
\end{eqnarray}
and
\begin{eqnarray}
H^k(s|d)\simeq\cases{H^k\left(\gamma|d,H_0(\delta)\right)\quad k\geq 0\cr
H_{-k}(\delta|d) \quad k<0\label{1.4}\cr}
\end{eqnarray}
(see \cite{HT} and I where this is recalled).
Here, $\gamma$ is the longitudinal
exterior derivative along the gauge orbits, denoted by $d$ (or $D$)
in \cite{HT}. The isomorphisms (\ref{1.1}) and (\ref{1.4})
are valid for arbitrary gauge theories and hold when the ``cochains"
(local $q$-forms) upon which $s$ acts are allowed to contain terms of
arbitrarily high antighost number.

Now, in the case of Yang-Mills models, the BRST differential is just the
sum of $\delta$ and $\gamma$,
\begin{equation}
s=\delta+\gamma
\end{equation}
and so, is {\it not} an infinite formal series of derivations
with arbitrarily high
antighost number (as it can a priori occur for an arbitrary gauge system).
It is thus natural to consider local $q$-forms that have bounded
antighost number, and to wonder whether the equalities
(\ref{1.1})-(\ref{1.4}) still hold under this restriction. Our first result,
derived in section \ref{3}, establishes precisely the validity of
(\ref{1.1})-(\ref{1.4}) in the space of local $q$-forms with bounded
antighost number.

The isomorphisms (\ref{1.1})-(\ref{1.4}) are useful in that they indicate
how BRST invariance is equivalent to - and can be used as a substitute
for - gauge invariance. However, they are not very explicit and a more
precise characterization of $H^k(s)$ or $H^k(s|d)$ is desired.

It has been shown in \cite{Henneaux1} that in each cohomological class of
$s$, one can find a representative that does not involve the antifields
and which is thus annihiliated by $\gamma$. It then easily follows that
\begin{equation}
H^k(s)\simeq H^k(\gamma,{\cal E})/{\cal N}\qquad (k>0)\label{1.6}
\end{equation}
where (i) ${\cal E}$ is the algebra generated by the vector potential
$A^a_\mu$, the ghosts $C^a$, the matter fields $y^i$ and their derivatives
(no antifields)~; and (ii) ${\cal N}$ is the ideal of elements of
${\cal E}$ that vanish on-shell. Since the cohomology of $\gamma$ in
${\cal E}$ is well understood in terms of Lie algebra cohomology, the
equation (\ref{1.6}) provides a more precise characterization of
$H^k(s)$ than (\ref{1.1}) does. The representatives of (\ref{1.6}) are
polynomials in the ``primitive forms" on the Lie algebra with
coefficients that are invariant polynomials in the field strengths, the matter
fields
and their covariant derivatives
\cite{Dixon,Bandelloni,Brandt1,Brandt,Brandt3,Violette1}.
Furthermore, two such objects are in the same class if they coincide
on-shell. To get a non redundant list, one may split the field strengths,
the matter fields and their covariant derivatives into ``independent"
components, which are not constrained by the equations of motion,
and ``dependent components", which may be expressed on-shell in terms of the
independent components. The cocyles may then
be chosen to depend only on the independent components. The isomorphism
(\ref{1.6}) is a cohomological reformulation of a theorem proved long
ago by Joglekar and Lee \cite{Joglekar}. It plays a crucial role in
renormalization theory \cite{Collins,Collins1}.

We derive in this paper an analogous, more precise characterization of the
local cohomology $H^k(s|d)$
of $s$ modulo $d$. For each value of the ghost degree,
and in arbitrary spacetime dimension,
we provide a constructive procedure for building representatives of each
cohomological class.  We then list all the solutions,
some of which are expressed in terms of non trivial
conserved currents which we assume to
have been determined. We find that contrary to what happens for the
cohomology of $s$, there exists cocycles in the cohomology of $s$
modulo $d$ from which the antifields cannot be eliminated by redefinitions.
Thus, there are new solutions to the consistency conditions
$sa+db=0$ besides the antifield independent ones, as pointed out in
\cite{Brandt2} for a Yang-Mills group with two abelian factors.

However, if the gauge group is semi-simple, these
additional solutions do not
arise at ghost number zero or one but only at higher ghost number. Accordingly,
the conjecture of Kluberg-Stern and Zuber on the renormalization
of (local and integrated)
gauge invariant operators \cite{Kluberg-Stern,Zinn-Justin1}
is valid in that
case (in even dimension).  Differently put, there
is no consistent perturbation of the Yang-Mills Lagrangian
of ghost number zero,
besides the perturbations by gauge invariant operators (or Chern-Simons
terms in odd dimensions).  Also, in four dimensions,
there is no new candidate gauge anomaly
besides the well known Adler-Bardeen one. Our results were partly
announced in \cite{BH} and do not use power counting.  They are purely
cohomological.

The BRST differential contains information about the dynamics of the
theory through the Koszul-Tate differential $\delta$.
Therefore, if
one replaces the Yang-Mills Lagrangian $-1/8 \ trF^2$ by a different
Lagrangian containing higher order derivatives of the
curvature, or Chern-Simons
terms in odd dimensions, the local BRST cohomology generically changes
even though the gauge transformations remain the same.  We show,
however, that the procedure for dealing with the Yang-Mills action
works also for these more general actions.

\section{BRST differential}\label{2}
\setcounter{equation}{0}
\setcounter{theorem}{0}

We assume throughout that the gauge group $G$ is compact and is thus
the direct product of a semi-simple compact group by abelian $U(1)$
factors. As in I, we take all differentials
to act from the right.

The BRST differential \cite{Becchi,Tyutin} for Yang-Mills models
is a sum of two pieces,
\begin{eqnarray}
s=\delta +\gamma,\ antigh\ \delta=-1,\ antigh\ \gamma=0
\end{eqnarray}
where $\delta$ is  explicitly given by
\begin{eqnarray}
\delta A^a_\mu=0,\ \delta C^a=0,\ \delta y^i=0\nonumber\\
\delta A^{*\mu}_a=-{\delta^L{\cal L}_0\over\delta A^a_\mu},
\ \delta C^*_a=-D_\mu A^{*\mu}_a + g T^j_{ai}y^*_j y^i,
\ \delta y^*_i= -{\delta^L{\cal L}_0\over\delta y^i}\label{KTYM}
\end{eqnarray}
Here, ${\cal L}_0={\cal L}^y_0 (y^i,D^y_\mu y^i)+{1\over 8}tr
F^{\mu\nu}F_{\mu\nu}$,
$D^y_\mu y^i=\partial_\mu y^i -g A^a_\mu T^i_{aj}y^j$, and
${\cal L}^y_0 (y^i,\partial_\mu y^i)$ is the free matter
field Lagrangian.
We assume for simplicity that the matter fields do not carry a
gauge invariance of their own and belong to a linear
representation of $G$. The differential $\gamma$ is given by
\begin{eqnarray}
\gamma A^a_\mu=D_\mu C^a,\ \gamma C^a=-{1\over 2}g C^a_{bc}C^b C^c,
\ \gamma y^i= g T^i_{aj}y^jC^a\nonumber\\
\gamma A^{*\mu}_a= g A^{*\mu}_c C^c_{ab}C^b,\ \gamma C^*_a=
 g C^*_c C^c_{ab}C^b,\nonumber\\ \gamma y^*_i= - g T^j_{ai}y^*_j C^a,
\label{GAYM}
\end{eqnarray}
There is no term of higher antighost number in $s$ because the
gauge algebra closes off-shell.
One has
\begin{eqnarray}
\delta^2=0,\ \gamma^2=0,\ \gamma\delta+\delta\gamma=0.
\end{eqnarray}

As explained in I, section 4, we shall consider local $q$-forms that are
polynomials in all the variables (Yang-Mills potential $A^a_\mu$,
matter fields $y^i$, ghosts $C^a$, antifields $A^{*\mu}_a$, $y^*_i$ and
$C^*_a$) and their derivatives. This is natural from the point of view
of quantum field theory and implies in particular that the local $q$-forms
under consideration have bounded antighost number.

Now, the general isomorphism theorems  (\ref{1.1})-(\ref{1.4})
have been established under the assumption that the local $q$-forms
may contain  terms of arbitrarily high antighost number. Our first
task is to refine the theorems to the
case where the allowed $q$-forms are constrained to have
bounded antighost number. This is done in the next section.

\section{Homological perturbation theory and bounded antighost number}
\label{3}
\setcounter{equation}{0}
\setcounter{theorem}{0}

\begin{theorem}{\em{\bf :}}\label{3.1}
for Yang-Mills models, the isomorphisms
\begin{eqnarray}
H^k(s)\simeq\cases{H^k\left(\gamma,H_0(\delta)\right)\quad  k\geq 0\cr
0 \quad k<0 \cr}
\end{eqnarray}
and
\begin{eqnarray}
H^k(s|d)\simeq\cases{H^k\left(\gamma|d,H_0(\delta)\right)\quad  k\geq 0\cr
H_{-k}(\delta|d) \quad k<0 \cr}\label{3.2}
\end{eqnarray}
also hold in the space of $q$-forms that are polynomials in all the variables
and their derivatives.
\end{theorem}

\proof{We extend the action of the even derivation $K$
of section 10 of I on the ghosts as follows,
\begin{eqnarray}
K=N_\partial+A
\end{eqnarray}
where $N_\partial$ is the operator counting the derivatives of all the
variables,
\begin{eqnarray}
N=\sum_{(k)} |k| \Bigl[{\partial^R\over\partial\left(\partial_{(k)}A^a_\mu
\right)}\partial_{(k)}A^a_\mu+{\partial^R\over\partial\left(\partial_{(k)}
C^a\right)} \partial_{(k)}C^a\nonumber\\
+{\partial^R\over\partial\left(\partial_{(k)}A^{*\mu}_a
\right)}\partial_{(k)}A^{*\mu}_a
+{\partial^R\over\partial\left(\partial_{(k)} C^*_a\right)}
\partial_{(k)}C^*_a\nonumber\\
+{\partial^R\over\partial\left(\partial_{(k)} y^i\right)}
\partial_{(k)}y^i+{\partial^R\over\partial\left(\partial_{(k)} y^*_i\right)}
\partial_{(k)}y^*_i\Bigr]
\end{eqnarray}
and where $A$ is defined by
\begin{eqnarray}
A=\sum_{(k)}\Bigl[2{\partial^R\over\partial\left(\partial_{(k)}A^{*\mu}_a
\right)}\partial_{(k)}A^{*\mu}_a+3{\partial^R\over\partial
\left(\partial_{(k)} C^*_a\right)} \partial_{(k)}C^*_a\nonumber\\
+2{\partial^R\over\partial\left(\partial_{(k)} \tilde y^*_i\right)}
\partial_{(k)}\tilde y^*_i+
{\partial^R\over\partial\left(\partial_{(k)} \bar y^*_i\right)}
\partial_{(k)}\bar y^*_i
-{\partial^R\over\partial\left(\partial_{(k)} C^a\right)}
\partial_{(k)}C^a\Bigr].\label{3.7}
\end{eqnarray}
The antifields $\tilde y^*_i$ are associated with second order differential
equations, while the antifields $\bar y^*_i$ are associated with first
order differential equations. We give $A$-weigth $-1$ to the ghosts
so that $\gamma$ has only components of non positive $K$-degree,
\begin{eqnarray}
\gamma =\gamma^0+\gamma^{-1},
\end{eqnarray}
just as $\delta$,
\begin{eqnarray}
\delta =\delta^0+\delta^{-1}+\delta^{-2}.
\end{eqnarray}
As shown in I, one has $[K,\partial_\mu]=\partial_\mu$ so that
the exterior derivative $d$ increases the eigenvalue of
$N_\partial$ and $K$ by one unit.

The ghosts are the only variables with negative $K$-degree ($\partial_\mu C^a$
has degree 0, $\partial_{\mu\nu} C^a$ has degree 1, etc\dots).
Furthermore, because the antifields carry all a strictly positive degree,
a form with bounded $K$-degree $k$ cannot contain terms of antighost number
greater than $k+g$, where $g$ is the dimension of the Lie algebra
(=number of ghosts). It is thus polynomial in the antifields.

We have indicated in section 10 of I that if $a$ is $\delta$-closed,
has positive antighost number and has $K$-degree bounded by $k$, then
$a=\delta b$ where $b$ has also $K$-degree bounded by $k$. Similarily,
if $a$ is $\delta$-closed modulo $d$, has both positive antighost
and pure ghost numbers, and has $K$-degree bounded by $k$, then
$a=\delta b +dc$ where $b$ has $K$-degree bounded by $k$ and $c$ has
$K$-degree bounded by $k-1$. Indeed, one knows from \cite{Henneaux2}
that $a=\delta b +dc$. The bound on the $k$-degree is then easily derived
by expanding the equality according to the $K$-degree, and using the
acyclicity of $\delta_0$, of $\delta_0$ mod $d$ and of $d$. These
properties are crucial in the proof of the theorem.

Let $a$ be a $s$-cocycle which is polynomial in all the variables and their
derivatives.
Let us expand $a$ according to the antighost number,
\begin{eqnarray}
a=a_0+a_1+\dots+a_m.
\end{eqnarray}
One has
\begin{eqnarray}
\delta a_{i+1}+\gamma a_i=0,\qquad i=0,1,2,\dots,m-1\label{3.12}
\end{eqnarray}
and
\begin{eqnarray}
\gamma a_m=0.
\end{eqnarray}
The isomorphism between $H^k(s)$ and $H^k\left(\gamma,H_0(\delta)\right)$
is defined by $[a]\mapsto [a_0]$.
To prove the theorem, one must verify that
this map is injective and surjective. This is done
as in \cite{HT}, by controlling further polynomiality through
the $K$-degree in a manner analogous
to what is done in I, section 10.  For instance, let us prove surjectivity.
Let $a_0$ be a representative of $H^k\left(\gamma,H_0(\delta)\right)$,
i.e., be an antifield independent solution of $\delta a_{1}+\gamma a_0=0$.
Since $a_0$ and $a_1$ are polynomials, they have bounded $K$-degree.  We
denote this bound by $k$.  To show that $a_0$ is the image of a polynomial
cocycle $a$ of $s$, one constructs recursively $a_2$, $a_3$ etc by means of
(\ref{3.12}). Because both $\delta$ and $\gamma$ have components of
non-negative $K$-degree, the higher order terms $a_2$, $a_3$ etc\dots
may be chosen to have
also $K$-degree bounded by $k$. Thus, the recursive construction stops
at antighost number $k+g$ (at the latest) and $a=a_0+a_1+\dots
+a_{k+g}$ is
polynomial. Injectivity, as well as (\ref{3.2}) are proved along the same
lines.}

To conclude, we note that theorem \ref{3.1} holds for all ``normal"
theories in the sense of section 10 of I, and, in particular,
for Einstein gravity. Moreover, the reader may check that that there is
some flexibility in the proof of the theorem, in that one may assign
different weights to the variables and nevertheless reach the same
conclusion.

\section{Cohomology of ${\bf \gamma}$}\label{4}
\setcounter{equation}{0}
\setcounter{theorem}{0}

In order to characterize completely $H^*(s|d)$, one needs a few preliminary
results. Some of them have been developed already in the literature,
while some of them are new. These results are: cohomology
$H^*(\gamma)$, invariant cohomology of $d$ and invariant cohomology of
$\delta$ modulo $d$.
They are considered in this section and the next two.

The cohomology $H^*(\gamma)$ of $\gamma$ has been computed
completely in
\cite{Dixon,Bandelloni,Brandt1,Brandt,Brandt3,Violette1,Henneaux1}.
The easiest way to describe it is to redefine the generators of the
algebra. The new generators
adapted to $\gamma$ are on the one hand $A^a_\mu$, its
symmetrized derivatives $\partial_{(\mu_1
\dots\mu_k}A^a_{\mu_{k+1})}$, ($k=1,2,\dots$) and their
$\gamma$-variations~; and on the other hand
$\chi^u_\Delta$ and the undifferentiated ghosts $C^a$, where the
$\chi^u_\Delta$ stand for the field strengths,
the matter fields, the antifields and all their covariant derivatives.
($u$ stands for representation indices~; while $\Delta$ stands for
spacetime or spinorial indices unrelated to the gauge group).
The $\chi^u_\Delta$
belong to a representation of the Lie algebra ${\cal G}$ of the gauge
group. Indeed, the field
strengths belong to the adjoint representation, the antifields
$A^{*\mu}_a$ and $C^*_a$
belong to the co-adjoint representation, while the antifields $y^*_i$
belong to the representation
dual to that of the $y^i$. As a result, the polynomials in the $\chi$'s
also form a representation
of the Lie algebra ${\cal G}$ of the gauge group: to any $x\in {\cal
G}$, there is a linear operator
$\rho (x)$ acting in the space of polynomials in the $\chi$'s as an
even derivation and such that
$\rho([x_1,x_2])=[\rho(x_1),\rho(x_2)]$. The representation $\rho$ is
completely reducible.
The polynomials belonging to the trivial representation are the
invariant polynomials.

The crucial feature in the calculation of $H^*(\gamma)$ is that
$A^a_\mu$, its symmetrized derivatives
and their $\gamma$-variations disappear from $H^*(\gamma)$ since
they belong to the ``contractible"
part of the algebra.
More precisely, one has

\begin{theorem}\label{12.1}{\em{\bf :}}
(i) The general solution of $\gamma a=0$ reads
\begin{eqnarray}
a=\bar a +\gamma b
\end{eqnarray}
where $\bar a$ is of the form
\begin{eqnarray}
\bar a =\sum \alpha_J(\chi^u_\Delta)\omega^J(C^a). \label{e12.4}
\end{eqnarray}
Here, the $\alpha_J$ are invariant polynomials in the $\chi$'s, while
the $\omega^J(C^a)$
belong to a basis of the Lie algebra cohomology of the Lie algebra of
the gauge group.

(ii) $\bar a$ is $\gamma$-exact if and only if
$\alpha_J(\chi^u_\Delta)=0$ for all $J$.
\end{theorem}

\proof{the proof may be found in
\cite{Dixon,Bandelloni,Brandt1,Brandt,Violette1,Henneaux1}
and will not be repeated here.}

Note that the $\alpha_J$ involve also the spacetime forms $dx^\mu$.
This will always be assumed in the sequel, where the word
``polynomial" will systematically mean ``spacetime
form with coefficients that are polynomial in the
variables and their derivatives".

\section{Invariant cohomology of ${\bf d}$}\label{5}
\setcounter{equation}{0}
\setcounter{theorem}{0}

Let $\alpha(\chi^u_\Delta)$ be an invariant polynomial in the
$\chi$'s. Assume that
$\alpha$ is $d$-closed, $d\alpha=0$. Then one knows from the theorem
on the cohomology of $d$
that $\alpha=d\beta$ for some $\beta$. Can one assume that $\beta$
is also an invariant polynomial?
If $\alpha$ does not contain the antifields, this may not be the case:
invariant polynomials
in the $2$-form $F^a\equiv (1/2)F^a_{\mu\nu}dx^\mu dx^\nu$ are
counterexamples (and the only ones)
\cite{Brandt,Violette1}. However, if $antigh\ \alpha >0$, one
has:

\begin{theorem}\label{d_gamma_cohomology}{\em{\bf :}}
the cohomology of $d$ in form degree $<n$ is trivial in the space of
invariant polynomials in the
$\chi$'s with strictly positive antighost number. That is, the
conditions
\begin{eqnarray}
\gamma\alpha=0,\ d\alpha=0,\ antigh\ \alpha>0,\ deg\ \alpha<n,\
\alpha=\alpha(\chi^u_\Delta)
\end{eqnarray}
imply
\begin{eqnarray}
\alpha =d\beta
\end{eqnarray}
for some invariant $\beta(\chi)$,
\begin{eqnarray}
\gamma \beta =0,
\end{eqnarray}
\end{theorem}

\proof{the proof proceeds as the proof of the proposition on page
363 in \cite{Violette1}.
We shall thus only sketch the salient points.

(i) First, one verifies the theorem in the abelian case with uncharged
matter fields. In that case,
any polynomial in the $\chi^u_\Delta$ is invariant since the $\chi$'s
themselves are invariant.
To prove the theorem in the abelian case, one splits the differential
$d$ as $d=d_0+d_1$,
where $d_1$ acts on the antifields only and $d_0$ on the other
fields. Let $\alpha$ be a polynomial
in the field strengths, the antifields, the matter fields and their
ordinary ($=$ covariant) derivatives.
If $d\alpha=0$, then $d_1\alpha^N=0$, where $\alpha^N$ is the piece
in $\alpha$ containing the
maximum number of derivatives of the antifields. But then,
$\alpha^N=d_1\beta^{N-1}$, where
$\beta^{N-1}$ is a polynomial in the $\chi^u_\Delta$. This implies
that $\alpha -d\beta^{N-1}$
ends at order $N-1$ rather than $N$. Going on in the same fashion,
one removes successively
$\alpha^{N-1},\alpha^{N-2},\dots$ until one reaches the desired
result.

(ii) Second, one observes that if $\alpha$ is invariant under a global
compact symmetry group,
then $\beta$ can be chosen to be also invariant since the action of
the group commutes with $d$.

(iii) Finally, one extends the result to the non-abelian case
with coloured matter fields by
expanding $\alpha$ according to
the number of derivatives of all the fields (see
\cite{Violette1} page 364 for the details).}

What replaces theorem \ref{d_gamma_cohomology} in form degree
$n$ is:
let $\alpha=\rho dx^0\dots dx^{n-1}$ be exact, $\alpha=d\beta$,
where $\rho$ is an invariant polynomial of antighost number
$>0$. [Equivalently, $\rho$ has vanishing variational derivatives with
respect to all the
fields and antifields]. Then, one may take the coefficients of the $(n-
1)$-form $\beta$
to be also invariant polynomials.

Theorem \ref{d_gamma_cohomology} can be generalized as follows.
Let $\alpha$ be a representative of
$H^*(\gamma)$, i.e.,
\begin{eqnarray}
\alpha=\Sigma \alpha_J(\chi^u_\Delta)\omega^J(C^a)
\end{eqnarray}
where the $\alpha(\chi)$ are invariant polynomials. Because
$d\gamma +\gamma d=0$, $d$ induces a well
defined differential in $H^*(\gamma)$. This may be seen directly as
follows. The derivative
$d\alpha_J=D\alpha_J$ is an invariant polynomial in the $\chi$'s
since $D$ commutes with the representation,
while $d\omega^J = \gamma\hat\omega^J(A,C)$ for
some $\hat\omega^J$. Thus $d\alpha= \pm
\Sigma(D\alpha_J)\omega^J+\gamma(\Sigma\alpha_J\hat\omega^J)$
defines an element of
$H^*(\gamma)$ ($\gamma\alpha_J=0$), namely the class of $\Sigma
D\alpha_J\omega^J\equiv \Sigma d\alpha_J\omega^J$. What is the
cohomology of $d$ in $H^*(\gamma)$? Again, we shall
only need the cohomology in form degree $<n$ and antighost number
$>0$.

\begin{theorem}\label{12.3}{\em{\bf :}}
$H^{g,l}_k(d,H^*(\gamma))=0$ for $k\geq1$ and $l<n$. Here $g$ is the
ghost number, $l$ is the
form degree and $k$ is the antighost number.
\end{theorem}

\proof{let $\alpha=\Sigma\alpha_J\omega^J$ be such that
$d\alpha=0$ in $H^*(\gamma)$, i.e.,
$d\alpha=\gamma\mu$. From the above calculation, it follows that
$\Sigma(D\alpha_J)\omega^J
=\gamma\mu^\prime$. But $\Sigma(D\alpha_J)\omega^J$ is of the
form (\ref{e12.4}). This implies that
$D\alpha_J=d\alpha_J=0$ by (ii) of theorem \ref{12.1}. Thus, by
theorem \ref{d_gamma_cohomology},
$\alpha_J=d\beta_J$ where $\beta_J$ are invariant polynomials in
the $\chi$'s. It follows that
$\alpha=\Sigma
d\beta_J\omega^J= \pm d(\Sigma\beta_J\omega^J)
\mp \gamma(\Sigma\beta_J\hat\omega^J)$ is indeed
$d$-trivial in $H^*(\gamma)$.}

Theorem \ref{12.3} is one of the main tools needed for the
calculation of $H^*(s|d)$
in Yang-Mills theory. It implies that there is no nontrivial
descent \cite{Stora,Zumino,Baulieu} for
$H(\gamma|d)$
in positive antighost number. Namely, if $\gamma a +db=0$, $antigh\
a>0$, one may redefine
$a\rightarrow a+\gamma\mu +d\nu=a^\prime$ so that $\gamma
a^\prime=0$. Indeed, the descent $\gamma a +db=0$,
$\gamma b +dc=0,\dots$ ends with $e$ so that $\gamma e=0$ and
$de+\gamma(something)=0$. Thus $e$ is
trivial and by the redefinition $e\rightarrow e +\gamma f +dm$, may
be taken to vanish, etc\dots .

\section{Invariant cohomology of ${\bf \delta}$ modulo ${\bf d}$}\label{6}
\setcounter{equation}{0}
\setcounter{theorem}{0}

The final tool needed in the calculation of $H^*(s|d)$ is the invariant
cohomology of
$\delta$ modulo $d$. We have seen that $H_k(\delta|d)$ vanishes for
$k>2$. Now, let
$\alpha$ be a $\delta$-boundary modulo $d$,
$\alpha=\delta\beta+d\gamma$, and let us assume that
$\alpha$ is an invariant polynomial in the $\chi$'s (no ghosts).
Can one also take
$\beta$ and $\gamma$ to be
invariant polynomials? The answer is affirmative as the next
theorem shows.

\begin{theorem}\label{12.4}{\em{\bf :}}
if the invariant polynomial $\alpha$ is a $\delta$-boundary modulo
$d$,
\begin{eqnarray}
\alpha=\delta\beta+d\gamma,
\end{eqnarray}
then one may assume that $\beta$ and $\gamma$ are also invariant
polynomials. In particular,
$H_k(\delta|d)=0$ for $k\geq3$ in the space of invariant
polynomials.
\end{theorem}

\proof{Let $a^k_p$ be a $k$-form of antighost number $p$ such that
\begin{eqnarray}
a^k_p=\delta\mu^k_{p+1} +d \mu^{k-1}_p,\qquad p\geq
1\label{13.1}.
\end{eqnarray}
We must show that both $\mu_{p+1}^k$ and $\mu_p^{k-1}$ may be
taken to be invariant
polynomials if $a^k_p$ is an invariant polynomial. To the equation
(\ref{13.1}),
we can associate a tower of equations that starts at form degree $n$
and ends at
form degree $k-p+1$ if $k\geq p$ or 0 if $k<p$,
\begin{eqnarray}
a^n_{p+n-k}=\delta\mu^n_{p+n-k+1} +d \mu^{n-1}_{p+n-k}\\
\vdots\nonumber\\
a^k_p=\delta\mu^k_{p+1} +d \mu^{k-1}_p\\
\vdots\nonumber\\
\cases{a^{k-p+1}_1=\delta\mu^{k-p+1}_2 +d \mu^{k-p}_1\cr
or\nonumber\cr
a^0_{p-k}=\delta\mu^0_{p-k+1},\cr}
\end{eqnarray}
where the $a$'s are all invariant polynomials. One goes up the ladder
by acting with $d$ and using
the fact that if an invariant polynomial is $\delta$-exact in the space
of all polynomials, then it is also
$\delta$-exact in the space of invariant polynomials (theorem 2 of
\cite{Henneaux1}). One goes down the
ladder by applying $\delta$ and using theorem
\ref{d_gamma_cohomology}.

It is easy to see, using again theorem 2 of \cite{Henneaux1} and
theorem \ref{d_gamma_cohomology}
that if any one of the $\mu_i^j$ is equal to an invariant polynomial
modulo $\delta$ or $d$
exact terms, then all of them fulfill that property. That is, if
$\mu^j_i=
M_i^j+\delta\rho_{i+1}^j+d\rho_i^{j-1}$ for one pair $(i,j)$ ($j-i=k-p-
1$), then
$\mu_l^m=M^m_l+\delta \rho^m_{l+1}+d\rho_l^{m-1}$ for all $(l,m)$.
Here, the $M^m_l$ are invariant
polynomials. Thus it suffices to verify the theorem for the top of the
ladder, i.e., the $n$-forms.
Furthermore, one has

\begin{lemma}{\em{\bf :}}\label{lem}
Theorem \ref{12.4} is obvious for $n$-forms of antighost number
$p>n$.
\end{lemma}

\proof{The proof is direct. If $a^n_p=\delta\mu_{p+1}^n+d\mu_p^{n-
1}$ with $p>n$, one gets at the
bottom of the ladder $a^0_{p-n}=\delta\mu_{p-n+1}^0$. But then, by
theorem 2 of \cite{Henneaux1},
one finds $\mu^0_{p-n+1} =M^0_{p-n+1}+\delta\rho^0_{p-n+2}$
where $M^0_{p-n+1}$ is an invariant
polynomial. This implies that all the $\mu$'s are of the required
form, and in particular
that $\mu_{p+1}^n$ and $\mu_p^{n-1}$ may be taken to be invariant
polynomials.}

We can now prove theorem \ref{12.4}. The proof proceeds as the
proof of theorem \ref{d_gamma_cohomology}.
Namely, one verifies first the theorem in the abelian case with a
single gauge field and uncharged matter fields. One then
extends it to the case of many abelian fields with a global symmetry.
One finally considers the full
non-Abelian case.

Since the last two steps are very similar to those of
theorem \ref{d_gamma_cohomology},
we shall only verify explicitly here that theorem \ref{12.4} holds for
a single abelian gauge field with uncharged matter fields.
So, let us start with a $n$-form $a_p$ solution of (\ref{13.1}) and
turn to dual notations,
\begin{eqnarray}
a_p=\delta b^\prime_{p+1}+\partial_\mu j^\mu_p\qquad (p\geq 1).
\end{eqnarray}
We shall first prove that if the theorem holds for antighost number
$p+2$, then it also holds for antighost
number $p$. A direct calculation yields
\begin{eqnarray}
{\delta a_p\over\delta C^*}=\delta Z^\prime_{(p-1)}\label{6.6}\\
{\delta a_p\over\delta A^{*\mu}}=\delta X^\prime_{(p)\mu}+
\partial_\mu Z^\prime_{(p-1)}\\
{\delta a_p\over\delta A_\mu}=\delta Y^{\prime\mu}_{(p+1)}-
\partial_\nu(\partial^\mu
X_{(p)}^{\prime\nu}-\partial^\nu X_{(p)}^{\prime\mu})\\
{\delta a_p\over\delta y^i}=D^+_{ji}X_{(p)}^{\prime i}+
\delta Y^\prime_{(p+1)i}\label{6.9}\\
{\delta a_p\over\delta y^*_i}=\delta X^{\prime i}_{(p)}\label{6.10}
\end{eqnarray}
where $Z^\prime_{p-1}$, $X^\prime_{(p)\mu}$,
$Y^{\prime\mu}_{p+1}$, $X_{(p)}^{\prime i}$ and $Y^\prime_{(p+1)i}$
are obtained by
differentiating $ b^\prime_{p+1}$ [$Z^\prime =0$ if $p=1$]. The
explicit expression of these
polynomials will not be needed in the sequel.
In (\ref{6.9}), $D^+_{ji}$ is the differential operator appearing in the
linearized matter equations of motion.
Because ${\delta^R
a_p/\delta C^*}$,
${\delta^R a_p/\delta A^{*\mu}}$, ${\delta^R a_p/\delta A_\mu}$,
$\delta^R a_p /\delta y^i$ and $\delta^R a_p /\delta y^*_i$ are
invariant polynomials, i.e., involve only the $\chi$'s, one may
replace in (\ref{6.6})-(\ref{6.10}) the polynomials
 $Z^\prime_{(p-1)}$, $X^\prime_{(p)\mu}$,
$Y^{\prime\mu}_{(p+1)}$, $X_{(p)}^{\prime i}$ and $Y^\prime_{(p+1)i}$
which may a priori
involve symmetrized derivatives of $A_\mu$, by
invariant polynomials
$Z_{(p-1)}$, $X_{(p)\mu}$,
$Y^{\mu}_{(p+1)}$, $X_{(p)}^{i}$ and $Y_{(p+1)i}$
depending only on the $\chi$'s,
\begin{eqnarray}
{\delta a_p\over\delta C^*}=\delta Z_{(p-1)}\label{z}\\
{\delta a_p\over\delta A^{*\mu}}=\delta X_{(p)\mu}-\partial_\mu
Z_{(p-1)}\label{xz}\\
{\delta a_p\over\delta A_\mu}=\delta Y^{\mu}_{(p+1)}-
\partial_\nu(\partial^\mu
X_{(p)}^{\nu}-\partial^\nu X_{(p)}^{\mu})\label{yx}\\
{\delta a_p\over\delta y^i}=D^+_{ji}X_{(p)}^{\prime i}+
\delta Y_{(p+1)i}\\
{\delta a_p\over\delta y^*_i}=\delta X^{i}_{(p)}.  \label{6.15}
\end{eqnarray}
This is obvious for $Z_{(p-1)}$ and $X^i_{(p)}$ (simply set
$A_\mu$ and its symmetrized derivatives equal to zero in
$Z^\prime_{(p-1)}$ and $X^\prime_{(p)}$; this commutes
with the action of $\delta$).  The assertion is then
verified easily for $X^\nu_{(p)}$, $Y_{(p+1)i}$ and
$Y^\mu_{(p+1)}$.

Now, the invariant polynomial $Y^\mu_{p+1}$ is $\delta$-closed
modulo $d$ by (\ref{yx}) since
${\delta a_p/\delta A_\mu}=\partial_\nu({\delta a_p/\delta
F_{\mu\nu}})$. Thus, it is $\delta$-exact
modulo $d$ because $H^{n-1}_{p+1}(\delta|d)\simeq H^n_{p+2}(\delta|d)$ is
empty ($p+2\geq 3$). This means
that $Y^\mu_{p+1}$ can be written as $\delta
T^\mu_{p+2}+\partial_{\nu}S^{\mu\nu}_{p+1}$ where
$T^\mu_{p+2}$ and $S^{\mu\nu}_{p+1}$ are both invariant
polynomials since we assume that the
theorems holds for antighost number $p+2$ in form degree $n$, or,
what is the same, by our
general discussion above, for antighost number $p+1$ in form degree
$n-1$.

If one injects relations (\ref{z}) - (\ref{6.15}) in the
identity
\begin{eqnarray}
a_p=\int\ dt[{\delta^R a_p\over\delta C^*}C^*+{\delta^R
a_p\over\delta A^{*\mu}}A^{*\mu}
+{\delta^R a_p\over\delta A_\mu}A_\mu+{\delta^R a_p\over\delta y^i}y^i
+{\delta^R a_p\over\delta y^*_i}y^*_i] +\partial_\mu
\rho^{\prime\mu}
\end{eqnarray}
one gets, using $Y^\mu_{p+1}=\delta
T^\mu_{p+2}+\partial_{\nu}S^{\mu\nu}_{p+1}$ and making
integrations by parts, that
\begin{eqnarray}
a_p=\delta b_{p+1}+\partial_\mu\rho^\mu
\end{eqnarray}
where $b_{p+1}$ is {\it manifestly invariant}. This proves that the
theorem holds in antighost
number $p$ if it holds in antighost number $p+2$ ($\rho^\mu$
may also be chosen to be
invariant by theorem \ref{d_gamma_cohomology}). But we know by
lemma \ref{lem} that the theorem is true
for antighost number $>n$. Accordingly, the theorem is true for all
(strictly) positive values of the antighost
number.}

\section{Calculation of ${\bf H^*(s|d)}$ - General method}\label{7}
\setcounter{equation}{0}
\setcounter{theorem}{0}

We can now turn to the calculation of $H^*(s|d)$ itself.
The strategy for computing $H^*(s|d)$ adopted here \cite{BH} is to relate
as much as possible elements of $H^*(s|d)$ to the known elements
of $H^*(\gamma|d)$
\cite{Violette2,Dixon,Bandelloni,Brandt1,Brandt,Brandt3,Violette1,Henneaux1}.
To that end, one controls the antifield dependence through theorems
\ref{12.3}
and \ref{12.4}. This is done by expanding the cocycle condition
$sa+db=0$ according to the antighost number.
At maximum antighost number $k$, one gets $\gamma a_k +
db_k=0$. Theorem \ref{12.3} and its
consequences for the descent equations for $\gamma$ in the
presence of antifields
then implies, for $k\geq 1$, that
one can choose $b_k$ equal to zero. Thus $\gamma a_k=0$, and by
theorem \ref{12.1}, $a_k=
\Sigma\alpha_{J}(\chi^u_\Delta)
\omega^J(C)$ up to $\gamma$-exact terms. [The redefinition
$a_k\rightarrow a_k +\gamma m_k +dn_k$
can be implemented through $a\rightarrow a+sm_k+dn_k$, which
does not change the class of $a$ in
$H(s|d)$]. The equation at antighost number $k-1$ reads $\delta a_k
+\gamma a_{k-1}
+db_{k-1}=0$. Acting with $\gamma$, we get $d\gamma b_{k-1}=0$,
which implies $\gamma b_{k-1}+
dc_{k-1}=0$.

If $k-1\geq 1$, theorem \ref{12.3} implies again that  one can choose
$\gamma b_{k-1}=0$ with
$b_{k-1}=\Sigma\beta_{J}(\chi^u_\Delta)\omega^J(C)$.
Inserting the forms of $a_k$
and $b_{k-1}$ into the equation
at antighost number $k-1$ gives
$\Sigma(\delta\alpha_{J}+d\beta_{J})\omega^J(C)
=\gamma(something)$ which implies $\delta \alpha_J + d\beta_J=0$
by part (ii) of theorem \ref{12.1}, i.e. $\alpha_J$ is
a $\delta$-cycle modulo $d$. Suppose that $\alpha_J$
is trivial, $\alpha_J=\delta\mu_J+d\nu_J$. Theorem \ref{12.4}
then implies that $\mu_J$
and $\nu_J$ can be choosen to be invariant polynomials. The
redefinition
$a\rightarrow a \pm s(\Sigma\mu_J\omega^J-
\Sigma\nu_J\hat\omega_J)-d(\Sigma \nu_J\omega^J)$
allows one to absorb $a_k$. [Recall that
$\gamma\hat\omega^J=d\omega^J$. The corresponding
redefinition of $b$ is $b\rightarrow b-s(\Sigma \nu_J\omega^J)$,
which leaves $b_k$ equal to zero
since $\gamma\nu_J=0$]. Consequently, we have learned (i)
that for $k \geq 1$, the last
term $a_k$ in any $s$-cocycle modulo $d$
may be chosen to be of the form $\Sigma\alpha_J\omega^J(C)$ where the
$\alpha_J$ are invariant polynomials~; and (ii)
that for $k \geq 2$, $\alpha_J$ define $\delta$-cycles
modulo $d$ which must be nontrivial since otherwise, $a_k$
can be removed from $a$ by adding to $a$ a $s$-
coboundary modulo $d$.

We can classify the elements of $H^*(s|d)$ according to their
last non trivial term in the antighost number expansion.
The results on the cohomology of $H_*(\delta|d)$ show that
only three cases are possible.

Class $I$: $a$ stops at antighost number 2,
\begin{eqnarray}
a=a_0+a_1+a_2\label{7.1}
\end{eqnarray}
(with $a_0=0$ if $gh\ a=-1$, or $a_0=a_1=0$ if $gh\ a=-2$). The last term
$a_2$ is invariant,
\begin{eqnarray}
a_2=\sum\alpha_J(\chi^u_\Delta)\omega^J(C)
\end{eqnarray}
and the $\alpha_J(\chi^u_\Delta)$ define non trivial elements
of $H_2(\delta|d)$.

Class $II$: $a$ stops at antighost number one,
\begin{eqnarray}
a=a_0+a_1\label{7.3}
\end{eqnarray}
(with $a_0=0$ if $gh\ a=-1$). The last term $a_1$ is invariant,
\begin{eqnarray}
a_1=\sum\alpha_J(\chi^u_\Delta)\omega^J(C)
\end{eqnarray}
We shall see in section 9 below
that the $\alpha_J(\chi^u_\Delta)$ must also
be non-trivial $\delta$-cycles modulo $d$.

Class $III$: $a$ does not contain the antifields,
\begin{eqnarray}
a=a_0.\label{7.5}
\end{eqnarray}
Then, of course, $gh\ a\geq 0$,

\section{Solutions of class $I$}\label{8}
\setcounter{equation}{0}
\setcounter{theorem}{0}

The solutions of class $I$ arise only when $H_2(\delta|d)$ is non trivial,
i.e., when there are free abelian gauge fields. This is a rather
academical context from the point of view of realistic Lagrangians,
but the question turns out to be of interest in the construction of
consistent couplings among free, massless vector particles
\cite{BH1}.

One can divide the solutions of class $I$ into three different types,
according to whether they have total ghost number equal to -2
(type $I_a$), -1 (type $I_b$) or $\geq 0$ (type $I_c$).

Type $I_a$: if $gh\ a=-2$, then $a$ reduces to $a_2$ and cannot involve the
ghosts. The solutions of type $I_a$ have form degree $n$ and are exhausted
by theorem 13.1 of I, in agreement with the isomorphism
$H^{-2}(s|d)\simeq H_2(\delta|d)$. They read explicitly
\begin{eqnarray}
a\equiv a_2=f^\alpha C^*_\alpha,\quad f^\alpha=constant
\end{eqnarray}
where $C^*_\alpha$ are the antifields conjugate to the ghosts of the
abelian, free, gauge fields. We switch back and forth between the form
notations and their dual notations. The $C^*_\alpha$
should thus be viewed alternatively
as $n$-forms or as densities.

Type $I_b$: if $gh\ a=-1$, then $a_2$ must involve one ghost $C^A$. This
ghost must be abelian since one must have $\gamma C^A=0$. Thus,
\begin{eqnarray}
a_2=f_{A\alpha}C^{*\alpha}C^A,\quad f_{A\alpha}=const.,
\end{eqnarray}
where the sum over $A$ runs a priori over {\it all} abelian ghosts. The
equation in antighost number one yields for $a_1$
\begin{eqnarray}
a_1=f_{A\alpha}A^A_\mu A^{*\alpha\mu}.
\end{eqnarray}
The next equation $\delta a_1+db_0=0$ is then equivalent to
\begin{eqnarray}
f_{A\alpha}F^A_{\mu\nu}F^{\alpha\mu\nu}=\partial_\rho k^\rho
\end{eqnarray}
for some $k^\rho$. This equality can hold only if the variational
derivatives of the left hand side identically vanishes, which implies
$f_{A\alpha}=0$ for $A\neq \beta$ and $f_{\alpha\beta}=-f_{\beta\alpha}$.
Thus, one gets finally
\begin{eqnarray}
a=f_{\alpha\beta}(A^\alpha_\mu A^{*\beta\mu}+C^\alpha C^{*\beta}),
\quad f_{\alpha\beta}=-f_{\beta\alpha}.
\end{eqnarray}

Type $I_c$: if $gh\ a\geq 0$, then all three terms $a_0$, $a_1$, and $a_2$
are in principle present. The term $a_2$ reads
\begin{eqnarray}
a_2=f_{\alpha J}C^{*\alpha}\omega^J(C)\label{8.6}
\end{eqnarray}
where $\omega^J(C)$ form a basis of the Lie algbra cohomology. The
$\omega^J(C)$ can be written as polynomials in the so-called ``primitive
forms". The primitive forms are of degree one $(C^A)$ for the
abelian factors and of degree $\geq 3$ ($trC^3$, $trC^5$, \dots) for each
simple factor \cite{Chevalley}.

It will be useful in the sequel to isolate explicitly the abelian ghosts in
(\ref{8.6}). Thus, we write
\begin{eqnarray}
a_2=\sum_k {1\over k!} f_{\alpha\Gamma A_1\dots A_k}
\omega^\Gamma (C)C^{A_1}\dots
C^{A_k}C^{*\alpha}\label{8.7}
\end{eqnarray}
where $\omega^\Gamma (C)$ involve only the ghosts of the simple factors.
The pure ghost numbers of the terms appearing in (\ref{8.7})
must of course add up to $2+q$, where $q$ is the total ghost number
of $a$. The factors $\omega^\Gamma (C)$ have the useful property of
belonging to a chain of descent equations \cite{Stora,Zumino,Baulieu}
involving at least two steps
\begin{eqnarray}
\partial_\mu \omega^\Gamma (C)=\gamma \hat \omega^\Gamma_\mu (C)\\
\partial_{[\mu}\omega^\Gamma_{\nu]} (C)=
\gamma \hat{\hat \omega}^\Gamma_{[\mu\nu]} (C)
\end{eqnarray}
For instance,
\begin{eqnarray}
\hat \omega^\Gamma_\mu = {\frac {\partial^R \omega^\Gamma}
{\partial C^a}} A^a_\mu
\end{eqnarray}
(see \cite{Violette2,Brandt}).  By contrast, the abelian ghosts
belong to a chain that stops after the first step.  One
has $\partial_\mu C^A = \gamma A^A_\mu$, but there is
clearly no $f_{\mu \nu}$ such that $\partial_{[\mu} A_{\nu]} =
\gamma f_{\mu \nu}$.  Since it will be necessary below to ``lift"
twice the elements $\omega^J(C)$ of the basis through
equations of the form (8.8) and (8.9), the abelian
factors play a distinguished role.

A direct calculation shows that
\begin{eqnarray}
\delta a_2= \gamma\Big[\Big(\sum {1\over (k-1)!}\omega^\Gamma
 f_{\alpha\Gamma A_1\dots A_k}C^{A_1}\dots
C^{A_{k-1}}A^{A_k}_\mu\nonumber\\
+\sum {1\over k!}(-)^k \hat \omega^\Gamma_\mu f_{\alpha\Gamma A_1\dots A_k}
C^{A_1}\dots C^{A_k}\Big)A^{*\alpha\mu}\Big]+\partial_\mu V^\mu
\end{eqnarray}
for some $V^\mu$. This fixes $a_1$ to be
\begin{eqnarray}
a_1=- \Big[\sum f_{\alpha\Gamma A_1\dots A_k}
\Big({1\over (k-1)!}\omega^\Gamma C^{A_1}\dots C^{A_{k-1}}
A^{A_k}_\mu\nonumber\\
+{1\over k!}(-)^k \hat \omega^\Gamma_\mu
C^{A_1}\dots C^{A_k}\Big)]A^{*\alpha\mu}\label{8.11}
\end{eqnarray}
up to a solution $m_1$ of $\gamma m_1 +dn_1=0$.
Using again the absence of non trivial descent in positive
antighost number, we may assume $n_1=0$ and $m_1=\sum_J
\mu_J(\chi^u_\Delta) \omega^J(C)$ by a redefinition $m_1\rightarrow
m_1+d\alpha+\gamma\beta$ that would only affect $a_0$ as $a_0\rightarrow
a_0+\delta\beta$ (if it exists). That is, $a_1$ takes the form
(\ref{8.11}) modulo an invariant object of antighost number one.

Compute now $\delta a_1$. One finds
\begin{eqnarray}
\delta a_1=- {1\over 2}\sum {1\over (k-1)!}\omega^\Gamma
 f_{\alpha\Gamma A_1\dots A_k}C^{A_1}\dots
C^{A_{k-1}}F^{A_k}_{\mu\nu}F^{\alpha\mu\nu}+\delta m_1\nonumber\\
+\gamma (M_{\mu\nu\alpha}F^{\alpha\mu\nu})
+\partial_\mu \tilde V^\mu\label{8.12}
\end{eqnarray}
for some $\tilde V^\mu$. Here, $M_{\mu\nu\alpha}$ is explicitly given by
\begin{eqnarray}
M_{\mu\nu\alpha}=\sum\Big[{1\over 2(k-2)!} f_{\alpha\Gamma A_1\dots A_k}
\omega^\Gamma C^{A_1}\dots
C^{A_{k-2}}A_\mu^{A_{k-1}}A_\nu^{A_k}\nonumber\\
+{2\over (k-1)!}(-)^k \hat \omega^\Gamma_{[\mu}f_{\alpha\Gamma A_1\dots A_k}
C^{A_1}\dots C^{A_{k-1}}A_{\nu]}^{A_k}\nonumber\\
-{1\over k!}\hat{\hat \omega}^\Gamma_{[\mu\nu]}f_{\alpha\Gamma A_1\dots A_k}
C^{A_1}\dots C^{A_k}\Big].
\end{eqnarray}
Thus, $\delta a_1$ is $\gamma$-closed modulo $d$ and $a_0$ exists if and
only if the first term on the right hand side of (\ref{8.12}) is weakly
$\gamma$-exact modulo $d$, i.e.,
\begin{eqnarray}
- {1\over 2}\sum {1\over (k-1)!}\omega^\Gamma
 f_{\alpha\Gamma A_1\dots A_k}C^{A_1}\dots
C^{A_{k-1}}F^{A_k}_{\mu\nu}F^{\alpha\mu\nu}+\delta m_1\nonumber\\
=\gamma m_0+\partial_\mu n_0^\mu\label{8.14}
\end{eqnarray}
for some $m_0$ and $n^\mu_0$ of antighost number zero. This forces this
first term to vanish, as we now show.

By acting with $\gamma$ on (\ref{8.14}), one gets $d\gamma n_0=0$ and
thus $\gamma n_0+dn^\prime_0=0$. Accordingly, $n_0$ is an antifield
independent solution of the $\gamma$-cocycle condition modulo $d$. This
equation has been completely solved in the literature
\cite{Dixon,Brandt,Brandt3,Violette1}
 and the solutions
fall into two classes: those that are annihilated by $\gamma$ and
are therefore invariant objects (up to redefinitions)~; and those that
lead to a non trivial descent, for which no redefinition can make
$n^\prime_0$ equal to zero. This second class involves only the forms
$A^a=A^a_\mu dx^\mu$,  $F^a=dA^a+ A^2$, their
exterior products, and the ghosts. Thus, $n_0=
\bar n_0 + \bar{\bar n}_0$, where $\bar n_0$ belongs to the first class and
$\bar{\bar n}_0$ belongs to the second class.

The solutions of the second class are obtained by considering the descent
$\gamma \bar{\bar n}_0+d \bar{\bar n}_0^\prime=0$,
$\gamma \bar{\bar n}_0^\prime+d \bar{\bar n}_0^{\prime\prime}=0$ etc \dots.
One successively lifts the last term of the descent, which is annihilated
by $\gamma$ all the way to $\bar{\bar n}_0$. The term $d\bar{\bar n}_0$
itself can be written as a $\gamma$-exact term, unless there is an
``obstruction". This obstruction is an invariant polynomial which involves
$\omega^J(C)$ and the components $F^{a}_{\mu\nu}$ but only through the forms
$F^a$ and their exterior products, but no other combination \cite{Violette2}.
In particular, the dual of $F^a$ cannot occur.
Accordingly, the obstruction cannot be written as a term involving
$F^A_{\mu\nu}F^{\alpha\mu\nu}$ plus a term involving the equations of
motion, plus a term of the form $d \bar n_0$, with $\bar n_0$
invariant. This means that the obstruction must be zero if
$a_0$ is to exist, so that
$d\bar{\bar n}_0=\gamma\mu_0$ by itself. By adding to
$a_0$ a solution of type $III$ if necessary,  we may assume $\bar{\bar n}_0$
to be absent.

If $n_0$ reduces to the invariant piece $\bar n_0$, the equation (\ref{8.14})
and theorem \ref{12.1} imply that
\begin{eqnarray}
-{1\over 2}\sum {1\over (k-1)!}\omega^\Gamma
 f_{\alpha\Gamma A_1\dots A_k}C^{A_1}\dots
C^{A_{k-1}}F^{A_k}_{\mu\nu}F^{\alpha\mu\nu}+\delta m_1
\nonumber\\- \sum (D_\mu n^\mu_J) \omega^J = 0
\end{eqnarray}
with $\bar n_0 = \sum n^\mu_J \omega^J$. If we set in this
equality the covariant
derivatives of $F^a_{\mu\nu}$ equal to zero, one gets the desired
result that  $ f_{\alpha\Gamma A_1\dots A_k}C^{A_1}\dots
C^{A_{k-1}}F^{A_k}_{\mu\nu}F^{\alpha\mu\nu}$ should vanish. This implies
that $ f_{\alpha\Gamma A_1\dots A_k}$ (i)
has as non vanishing components only
$f_{\alpha\Gamma\alpha_1\dots\alpha_k}$; and (ii) is completely
antisymmetric in ($\alpha,\alpha_1,\dots,\alpha_k$). The solutions of class
$I_c$ are consequently exhausted by
\begin{eqnarray}
a=\sum f_{\alpha\Gamma\alpha_1\dots\alpha_k}\Big[
\Big(-{1\over 2(k-2)!}\omega^\Gamma C^{\alpha_1}\dots
C^{\alpha_{k-2}}A_\mu^{\alpha_{k-1}}A_\nu^{\alpha_k}\nonumber\\
-{2\over (k-1)!}(-)^k \hat \omega^\Gamma_{[\mu}
C^{\alpha_1}\dots C^{\alpha_{k-1}}A_{\nu]}^{\alpha_k}
+{1\over k!}\hat{\hat \omega}^\Gamma_{[\mu\nu]}
C^{\alpha_1}\dots C^{\alpha_k}\Big)
F^{\alpha\mu\nu}\nonumber\\
\nonumber\\
\nonumber\\
-\Big({1\over (k-1)!}\omega^\Gamma C^{\alpha_1}\dots C^{\alpha_{k-1}}A^{A_k}+
{1\over k!}(-)^k \hat \omega^\Gamma_\mu
C^{\alpha_1}\dots C^{\alpha_k}\Big)A^{*\alpha\mu}\nonumber\\
+{1\over k!}\omega^\Gamma C^{\alpha_1}\dots C^{\alpha_k}C^{*\alpha}\Big]
\end{eqnarray}
(modulo solutions of class $II$).
This ends our discussion of the solutions of class I, corresponding to
elements of $H_2(\delta|d)$.

[The analysis has been performed explicitly for spacetime dimensions
greater than or equal to three.  In two spacetime dimensions, there
are further solutions.  The solutions of ghost number $-2$ read
$(\partial f/\partial F^a_{01})  C^*_a + (1/2) (\partial^2 f /
\partial F^b_{01} \partial F^a_{01}) \epsilon_{\mu \nu} A^{*\mu}_a A^{*\nu}_b$,
where $f$ is an invariant polynomial in those field strengths $F^a_{\mu \nu}$
that obey $D_\mu F^a_{01} = 0$ on-shell.  The solutions of ghost
number $-1$ and higher are constructed as above, by multiplying the
solutions of ghost number $-2$ with the $\gamma$-invariant polynomials
$\omega^J (C)$, and then solving successively for $a_1$ and $a_0$.
There are possible obstructions in the presence of abelian factors
which restrict the coefficients
of $\omega^J$.  We leave the details to the reader.]

\section{Solutions of class $II$}\label{9}
\setcounter{equation}{0}
\setcounter{theorem}{0}

The next case to consider is given by a cocycle $a$ whose expansion stops
at antighost number 1. Again, we may consider two subcases: type $II_a$, with
$gh\ a=-1$~; and type $II_b$, with $gh\ a\geq 0$.

Type $II_a$: if $gh\ a=-1$, then $a$ reduces to $a_1$ and does not
involve the ghosts. It is clearly an element of $H_1(\delta|d)$,
by the equation $\delta a_1 + d b_0 = 0$. The groups
$H_1^k(\delta|d)$ are non empty in form degree $n$ (conserved currents)
and $n-1$ (if there are uncoupled abelian fields). Let $j^\mu_\Delta$ be
a complete set of inequivalent non trivial conserved currents and let
$X^a_{\mu\Delta}$, $X^i_\Delta$ be the corresponding global symmetries
of the fields,
\begin{eqnarray}
\delta(X^a_{\mu\Delta}A^{*\mu}_a+X^i_\Delta y^*_i)=\partial_\mu j^\mu_\Delta
\end{eqnarray}
We impose to $X^a_{\mu\Delta}A^{*\mu}_a+X^i_\Delta y^*_i$ to be annihilated
by $\gamma$ i.e., to be invariant. Because the equations of motion
involve derivatives of the field strengths, and are not invariant
polynomials in the forms $F^a$, there is no obstruction to taking
$j^\mu_\Delta$ annihilated by $\gamma$ as well.

One gets for the BRST cohomology $H^{-1}(s|d)$.

In form degree $n-1$:
\begin{eqnarray}
a=f^\alpha A^{*\mu}_\alpha,\quad f^\alpha=constant.
\end{eqnarray}

In form degree $n$:
\begin{eqnarray}
a=f^\Delta(X^a_{\mu\Delta}A^{*\mu}_a+X^i_\Delta y^*_i),\quad
f^\Delta=constant.
\end{eqnarray}
Turn now to the solutions of type $II_b$.

Type $II_b$: We must solve $\gamma a_0 +
\delta a_1 + db_0 = 0$ with $ a_1 = \sum \alpha_J
\omega^J$.  The derivation above does not imply that
$b_0$ is annihilated by $\gamma$ and thus,
it is not clear at this stage that the $\alpha_J$ belong to
$H(\delta|d)$.  However, by acting with
$\gamma$ on
$\delta a_1 + \gamma a_0+db_0=0$, one gets again that $\gamma b_0 + dc_0=0$.
The analysis proceeds then in a manner similar
to that of type $II_c$.  As mentioned above, the
general
solution to $\gamma b_0 + dc_0 = 0$ is
known \cite{Dixon,Brandt,Brandt3,Violette1} and
takes the form
$b_0=\bar b_0 +
{\bar{\bar b}}_0$, where (i) $\bar b_0$ is annihilated by
$\gamma$ and thus given by $\bar
b_0=\Sigma\beta_{0J}(\chi)\omega^J(C)$
(up to irrelevant $\gamma$-exact terms) with $\beta_{0J}$
invariant polynomials in the $\chi$'s~;
and (ii) ${\bar{\bar b}}_0$ is
$\gamma$ closed only modulo a non-trivial $d$ exact term
and involves the forms $A^a=A^a_\mu dx^\mu$,
$F^a= dA^a+ A^2$, and $C^a$.
The obstruction \cite{Violette2} to writing $d{\bar{\bar
b}}_0$ as a
$\gamma$ exact term involves
the forms $F^a$ and $\omega^J(C)$.  It cannot be written as  the
sum of a
term proportional to the equations of motion and a term of the form
$d\bar b_0$
with $\bar b_0$ invariant since such terms involve unavoidably
the covariant derivatives of the field strengths.
Thus, the obstruction must be absent and $d {\bar{\bar b}}_0 =
- \gamma {\bar {\bar a}}_0$, for some ${\bar {\bar a}}_0$.
The equation $\delta a_1 +\gamma a_0 + db_0 = 0$ splits therefore into two
separate
equations $\gamma {\bar{\bar a}}_0+d{\bar{\bar
b}}_0=0$ and
$\gamma \bar a_0 +d\bar b_0 +\delta a_1=0$.

The first equation defines a solution of class III.
We need only consider in this section
the second equation.  Because $\bar b_0$ is annihilated by $\gamma$, we may
follow the procedure of section 7 to find again that
the invariant polynomials $\alpha_J$ in $a_1$ define elements
of $H_1(\delta|d)$.
One gets explicitly.

In form degree $n-1$:
\begin{eqnarray}
a=f^\alpha_J ( {\hat \omega}^J_\nu (C)F^{\mu \nu}_\alpha +
\omega^J(C) A^{*\mu}_\alpha),\quad f^\alpha_J=constant.
\end{eqnarray}

In form degree $n$:
\begin{eqnarray}
a=f^\Delta_J[{\hat \omega}^J_\mu j^\mu_\Delta +
\omega^J (C)(X^a_{\mu\Delta}A^{*\mu}_a+X^i_\Delta y^*_i)],\quad
f^\Delta_J=constant.
\end{eqnarray}
(with $\gamma {\hat \omega}^J_\mu = \partial_\mu \omega^J$).

[In two dimensions, there are further solutions obtained by
taking $f^a_J = \partial f_J /\partial F^a_{01}$, where $f_J$ are arbitrary
invariant polynomials in the $F^a_{01}$.  We leave the details to
the reader.]

The solutions of class $I$ exist only if there are free,
abelian gauge fields.  For a semi-simple gauge group,
class $I$ is empty.  By contrast, the solutions
of class $II$ in form degree $n$ exist whenever there are
non trivial conserved currents,
or, equivalently, non trivial global
symmetries.  They occur at ghost number $-1$, or $-1 + l_J$, where
$l_J$ is the ghost number of the element $\omega^J$ of the
chosen basis of the
Lie algebra cohomology.  For a semi-simple gauge group,
$l_J$ is greater than or equal to three.  Thus, the solutions
of class $II$ occur at ghost number $-1$, $2$, and higher, but
not at ghost number $0$ or $1$.  The solutions at ghost number
$2$ are given by (9.5) with $\omega^J =$ tr$C^3$ and
$\hat \omega^J_\mu = 3$tr$C^2 A_\mu$.

We close this section by pointing out that one may regroup
the conserved currents $j_\Delta$ (viewed as ($n-1$)-forms)
and the coefficients $X^i_\Delta$
into a single object
\begin{eqnarray}
\bar G_\Delta = d^n x (X^a_{\mu\Delta}A^{*\mu}_a+X^i_\Delta y^*_i ) + j_\Delta,
\end{eqnarray}
which has the remarkable property of being annihilated
by  the sum $\bar s = s +d$,
\begin{eqnarray}
\bar s \bar G_\Delta = 0.
\end{eqnarray}
This equation is the analog of a similar equation holding for $\bar
q^*_\alpha$,
\begin{eqnarray}
\bar q^*_\alpha = C^*_\alpha + A^*_\alpha + *F_\alpha
\end{eqnarray}
where the $C^*_\alpha$ are viewed as $n$-forms, the $A^*_\alpha$ are
viewed as $(n-1)$-forms
and the dual $*F_\alpha$ to the uncoupled free abelian field strength
are $(n-2)$-forms.  One has also
\begin{eqnarray}
\bar s q^*_\alpha = 0.
\end{eqnarray}
In veryfing these relations, one must use explicitly the fact that the
spacetime dimension is $n$ through $d$($n$-form) $=0$.

\vfill
\eject

\section{Non-triviality of solutions of classes $I$ and $II$}\label{nontri}
\setcounter{equation}{0}
\setcounter{theorem}{0}

We verify in this section that the solutions of types
$I$ and $II$ are all non trivial.

\begin{theorem}{\em{\bf :}}
any BRST cocycle $a$ modulo $d$ belonging to the class I or to the class II
is necessarily non trivial, $a\neq sc+de$.
\end{theorem}

\proof{the idea of the proof is to show that if $a=sc+de$, then, the
$\alpha_J(\chi^u_\Delta)$ all define trivial elements of $H_2(\delta|d)$
or $H_1(\delta|d)$. So, assume that $a=sc+de$. Expand this equation
according to the antighost number. One gets
\begin{eqnarray}
a_0=\gamma c_0+\delta c_1 +de_0,\quad a_1=\gamma c_1+\delta c_2 +de_1
\end{eqnarray}
and
\begin{eqnarray}
0=\gamma c_i +\delta c_{i+1} +d e_i\qquad (i\geq2)\label{7.7}
\end{eqnarray}
(we assume $a$ to belong to the class $II$ for definiteness~;
the argument
proceeds in the same way for the class $I$). Let $c$ stop at antighost
number $M$, $c=c_0+c_1+\dots+c_M$. Then, one may assume that $e$
stops also at antighost number $M$. Indeed, the higher order terms can be
removed from $e$
by adding a $d$-exact term since $H^k(d)=0$ for $k<n-1$.
Now, the equation (\ref{7.7}) for $i=M$ reads $\gamma c_M+de_M=0$ and
is precisely of the form analysed above. Since $M\geq 2$, one may assume
$e_M=0$ and then by adding to $c_M$ a $s$-exact modulo $d$-term
(which does not modify $a$), that $c_M$ is of the form
$c_M=\sum\gamma_J(\chi^u_\Delta)\omega^J(C)$. Next, the equation at order
$M-1$ shows that $c_M$ can actually be removed, unless $M=2$. Thus,
we may assume without loss of generality that $c=c_0+c_1+c_2$,
$c_2=\sum\gamma_J(\chi^u_\Delta)\omega^J(C)$ and $e=e_0+e_1$. It follows
that the equation for $a_1$ reads
\begin{eqnarray}
\sum\alpha_J(\chi^u_\Delta)\omega^J(C)=\gamma c_1+\sum\delta
\gamma_J(\chi^u_\Delta)\omega^J(C)+de_1.\label{7.8}
\end{eqnarray}
By acting with $\gamma$ on this equation, we obtain as above that
$e_1$ may also be chosen to be invariant,
$e_1=\sum\varepsilon_J(\chi^u_\Delta)\omega^J(C)$.
Accordingly, (\ref{7.8}) reads
\begin{eqnarray}
\sum\left(\alpha_J(\chi^u_\Delta)-\delta\gamma_J(\chi^u_\Delta)-
d\varepsilon_J(\chi^u_\Delta)\right)\omega^J(C)=\gamma c^\prime_1
\end{eqnarray}
from which one infers, using theorem \ref{12.1}, that
\begin{eqnarray}
\alpha_J(\chi^u_\Delta)-\delta\gamma_J(\chi^u_\Delta)-
d\varepsilon_J(\chi^u_\Delta)=0.
\end{eqnarray}
This shows that all the $\alpha_J$ are $\delta$-exact modulo $d$,
in contradiction to the fact that they define non trivial
elements of $H_*(\delta|d)$. Therefore, the cocycle $a$ cannot be $s$-exact
modulo $d$.}

\section{Solutions of class $III$}\label{10}
\setcounter{equation}{0}
\setcounter{theorem}{0}

The solutions of class $III$ do not depend
on the antifields and fulfill $\gamma a_0+db_0
=0$.  As we have recalled, these equations have been
extensively studied previously  and their
general solution is known
\cite{Dixon,Violette2,Brandt,Brandt3,Violette1}.
For this reason, we refer the reader
to the existing literature for their explicit construction.

The solutions are classified according to whether $b_0$ can be
removed by redefinitions or not.

Type $III_a$ :  $\gamma a_0 = 0$.
The abelian anomaly $C F_{\mu \nu} dx^\mu dx^\nu$ in two dimensions
belongs to this class.

Type $III_b$ : $\gamma a_0+db_0
=0$, with $b_0$ non trivial.  In that case, $a_0$ and $b_0$ may be assumed to
depend only on the forms $A^a$, $F^a$,  $C^a$ and their exterior products.

The elements of $H(\gamma|d)$ not involving the antifields are
non trivial as elements of $H(s|d)$ if and only if they do not vanish
on-shell modulo $d$.  Thus, the non trivial elements of
$H(\gamma|d)$ of type $III_b$ remain non trivial as elements of $H(s|d)$ since
the
forms $A^a$ and $F^a$ are unrestricted by the equations of motion.
However, the solutions of type $III_a$ may become trivial even if they
are non trivial as elements of $H(\gamma|d)$.

The solutions of direct interest are those of ghost number
zero and one.  At ghost number zero, class $III_a$ contains the
invariant polynomials in the field strengths, the matter fields
and their covariant derivatives.  The Yang-Mills Lagrangian
belongs to class $III_a$.  Class $III_b$ contains non trivial
solutions at ghost number zero only in odd
spacetime dimensions $2k+1$.
These non trivial solutions are the
Chern-Simons terms, given by
\begin{eqnarray}
{\cal L}_{CS} = tr(A F^k + \dots)
\end{eqnarray}
where the dots denote polynomials in
$A^a$ and $F^a$ whose degree in $F$ is smaller than $k$ and whose
form degree equals $2k +1$.

At ghost number one, type $III_a$ contains solutions
of the form ``abelian ghost times invariant polynomial".
It contains no solution if the group is semi-simple.
Type $III_b$ contains the famous Adler-Bardeen anomaly.

\section{More general Lagrangians}\label{13}
\setcounter{equation}{0}
\setcounter{theorem}{0}

In the previous discussion, we have assumed that the Lagrangian was
the standard Yang-Mills Lagrangian.  This assumption was explicitly
used in the calculation since the dynamics enters the BRST differential
through the Koszul-Tate differential.

It turns out, however, that for a large class of Lagrangians, one can
repeat the analysis and get similar conclusions.
These Lagrangians are gauge invariant up to a total derivative
and thus read
\begin{eqnarray}
{\cal L} = {\cal L}_0 (y,F_{\mu \nu}, D_\mu y, D_\rho F_{\mu \nu},\dots)
+ {\cal L}_{CS}
\end{eqnarray}
where ${\cal L}_0$ is an invariant polynomial in the matter fields,
the fields strengths and their covariant derivatives, and where the
Chern-Simons
term ${\cal L}_{CS}$ is available only in odd dimensions.
We shall assume that the Yang-Mills gauge symmetry exhausts all
the gauge symmetries.  We shall also
impose that the Lagrangian ${\cal L}$ defines a normal
theory in the sense of section 10 of I. The calculation
of $H(s|d)$ can then be performed along the lines of this paper.

(i) First, one verifies that the $\gamma$-invariant
cohomology $H_k(\delta |d)$ is described
as before : $H_k(\delta |d)$ vanishes for $k$ strictly
greater than $2$; for $k=2$, it is
non-empty only if there are uncoupled
abelian gauge fields, in which case it is spanned by $C^*_\alpha$;
and for $k=1$, it is isomorphic to the set of non trivial
global symmetries with invariant $a_1$.
Thus, the dynamics enters explicitly
 $H_k(\delta |d)$ only at $k=1$, through the conserved currents.

(ii) The solutions of class $I$ makes a further use of the dynamics through
the study of the obstructions to the existence of $a_0$.  A case by case
analysis, which proceeds as in section 8, is in principle required.
Recall, however, that class $I$ exists only in the
academic situation where there are uncoupled abelian gauge fields.

(iii) Class $II$ also uses the equations of
motion in the proof that $a_1$ should define elements of
$H(\delta|d)$.  It must be verified whether the equations
of motion can or cannot remove obstructions given by polynomials
in the forms $F^a$.  Again, the analysis proceeds straightforwardly
as in section 9.

(iv) Class $III$ is obviously unchanged since it
does not involve the antifields (only the
coboundary condition is modified, since the concept of ``on-shell
trivial" is changed).

The analysis is particularly simple for the pure Chern-Simons theory
in three dimensions, without the Yang-Mills part.  We take a semi-simple
gauge group.  Class $I$ is then empty.  Class $II$ is empty as well since
there is no non-trivial $a_1$ annihilated by $\gamma$.  Only
class $III$ is present. Among the solutions of class $III$, those that
are of the subtype $III_a$ turn out to be trivial since the field strengths
and their covariant derivatives vanish on-shell.  Thus, we are left
with class $III_b$.  These solutions are obtained from the standard
descent, with bottom given by the elements $\omega^J$ of the basis
of the Lie algebra cohomology ($tr C^3$, $trC^5$ etc), with constant
coefficients (no $F$ since $F=0$ on-shell).  This agrees with
the analysis of \cite{Piguet}.

\section{Conclusion}\label{14}
\setcounter{equation}{0}
\setcounter{theorem}{0}

In this paper, we have explicitly computed the
cohomological groups $H^k(s|d)$
for Yang-Mills theory.
Our work goes beyond previous analyses on the subject
\cite{Becchi,Stora,Dixon,BBlasi,Bonora,BTM,TM,Violette2,Baulieu,Bandelloni,Brandt,Violette1},
in that (i) we do not
use power counting; and (ii) we explicitly include the
antifields (= sources for the BRST variations). We
have shown that
new cohomological classes depending on the antifields appear
whenever there are conserved
currents, but only at antighost number $\geq 2$ for a semi-simple
gauge group.
Our results confirm previous conjectures in the field.
[The existence of antifield-dependent
solutions of the consistency equation
at ghost number one for a theory
with abelian factors was anticipated in
\cite {BBlasi}.  The structure of these solutions
was partly elucidated
and an argument was given that they cannot occur as anomalies].

The central feature behind our analysis
is a key property of the antifield formalism, namely,
that the antifields provide a resolution of the stationary surface
through the Koszul-Tate differential \cite{HT}.  It is by attacking
the problem from that angle that we have been able to carry
out the calculation to completion, while previous attempts
following different approaches turned out to be unsuccessful.
Thus, even in the familiar
Yang-Mills context, the formal ideas of the antifield formalism
prove to be extremely fruitful.

Our results can be extended in various directions. First, one can
repeat the
Yang-Mills calculation for Einstein gravity with or without matter.
This will be
done explicitly
in \cite{BBH1}. Second, at a more theoretical level, one can analyze
further the connection between the local BRST cohomology,
the characteristic cohomology and the variational bicomplex \cite{Ander}.
This will be pursued in
\cite{BBH2}.

\section{Acknowledgements}

We are grateful to J. Collins, M. Dubois-Violette, O. Piguet, A. Slavnov, J.
Stasheff,
R. Stora, M. Talon, C. Teitelboim and C. Viallet for fruitful discussions.
This work has been supported in part by research funds from the
Belgian ``F.N.R.S."
as well as by research contracts with the Commission of the European
Communities. M.H. is grateful to the CERN theory division for
its kind hospitality while this work was being
carried out and completed.

\vfill
\eject


\begin{thebibliography}{999}
\baselineskip=12truept
\bibitem{BBH} G. Barnich, F. Brandt and M. Henneaux,
{\em Local BRST cohomology in the antifield formalism: I. General theorems},
preprint ULB-TH-94/06, NIKHEF-H 94-13, hep-th 9405109.
\bibitem{HT} M. Henneaux and C. Teitelboim, {\em
Quantization of
Gauge Systems}, Princeton University Press (Princeton: 1992).
\bibitem{Henneaux1} M. Henneaux, {\em Phys. Lett.}
{\bf B 313} (1993) 35.
\bibitem{Dixon} J.A. Dixon, {\em Cohomology and Renormalization
of Gauge Theories I, II, III}, unpublished preprints (1976-1979)~;
{\em Commun. Math. Phys.} {\bf 139} (1991) 495.
\bibitem{Bandelloni} G. Bandelloni, {\em J. Math. Phys.} {\bf 27}
(1986) 2551~;
{\bf 28} (1987) 2775.
\bibitem{Brandt1} F. Brandt, N. Dragon and M. Kreuzer,
{\em Phys. Lett.} {\bf B231} (1989) 263.
\bibitem{Brandt} F. Brandt, N. Dragon and M. Kreuzer,
{\em Nucl. Phys.} {\bf B332} (1990) 224.
\bibitem{Brandt3} F. Brandt, N. Dragon and M. Kreuzer,
{\em Nucl. Phys.} {\bf B332} (1990) 250.
\bibitem{Violette1} M. Dubois-Violette, M. Henneaux, M.
Talon
and C.M. Viallet, {\em Phys. Lett.} {\bf B 289} (1992) 361.
\bibitem{Joglekar} S.D. Joglekar and B.W. Lee, {\em Ann. Phys.} (NY)
{\bf 97} (1976) 160.
\bibitem{Collins} J.C. Collins, {\em Renormalization} (Cambridge U.P.,
Cambridge 1984).
\bibitem{Collins1} J.C. Collins and R.J. Scalise {\em The
renormalization of composite operators in Yang-Mills
theories using general covariant gauge}, preprint PSU/TH/141, hep-
ph/9403231.
\bibitem{Brandt2} F. Brandt, {\em Phys. Lett.} {\bf B320} (1994) 57.
\bibitem{Kluberg-Stern}  H. Kluberg-Stern and J.B. Zuber,
{\em Phys. Rev.} {\bf D12} (1975) 467, 482, 3159.
\bibitem{Zinn-Justin1} J. Zinn-Justin, {\em Quantum Field Theory and
Critical Phenomena}, $2^{\rm nd}$ Edition, Clarendon Press
(Oxford: 1993).
\bibitem{BH} G. Barnich and M. Henneaux, {Phys. Rev. Lett.}
{\bf 72} (1994) 1588.
\eject
\bibitem{Becchi} C. Becchi, A. Rouet and R. Stora, {\em Commun.
Math. Phys.} {\bf 42} (1975) 127~; {\em Ann. Phys.} (N.Y.) {\bf 98}
(1976) 287.
\bibitem{Tyutin} I.V. Tyutin, {\em Gauge Invariance in Field Theory and
Statistical Mechanics}, Lebedev preprint FIAN n$^0$ 39 (1975).
\bibitem{Henneaux2} M. Henneaux, {\em Commun. Math. Phys.} {\bf
140} (1991) 1.
\bibitem{Stora} R. Stora, {\em Continuum Gauge Theories} in
{\em New Developments in Quantum Field Theory and Statistical Mechanics},
ed. M. L\'{e}vy and P. Mitter (Plenum: 1977), {\em Algebraic
Structure and Topological Origin of Anomalies} in {\em Progress in
Gauge Field Theory}, ed. G. 't Hooft et al. (Plenum: 1984).
\bibitem{Zumino} B. Zumino, {\em Chiral
Anomalies and Differential Geometry} in
{\em Relativity, Groups and Topology II}, ed. B.S. De Witt and R. Stora
(North-Holland: 1984)
\bibitem{Baulieu} L. Baulieu, {\em Phys. Rep.} {\bf 129} (1985) 1.
\bibitem{Violette2} M. Dubois-Violette, M. Talon and C.M.
Viallet, {\em Phys. Lett.} {\bf B158} (1985) 231~;
{\em Commun. Math. Phys.} {\bf 102} (1985) 105.
\bibitem{BH1} G. Barnich and M. Henneaux, {\em Phys. Lett.}
{\bf B311} (1993) 123.
\bibitem{Chevalley} C. Chevalley and S. Eilenberg, {\em Trans.
Am. Math. Soc.} {\bf 63} (1953) 589~; J.L. Koszul, {\em Bull.
Soc. Math. France} {\bf 78} (1950) 65~; G. Hochschild and J.P. Serre,
{\em Ann. Math.} {\bf 57} (1953) 591.
\bibitem{Piguet} F. Delduc, C. Lucchesi, O. Piguet and S.P. Sorella,
{\em Nucl. Phys.} {\bf B346} (1990) 313~; A. Blasi, O. Piguet and
S.P. Sorella, {\em Nucl. Phys.} {\bf B356} (1991) 154~; C. Lucchesi
and O. Piguet, {\em Nucl. Phys.} {\bf B381} (1992) 281.
\bibitem{BBlasi} G. Bandelloni, A. Blasi, C. Becchi and
R. Collina, {\em Ann. Inst. Henri Poincar\'e}, {\bf 28} (1978)
225, 255.
\bibitem{Bonora} L. Bonora and P. Cotta-Ramusino, {\em Commun. Math. Phys.}
{\bf 87} (1983) 589.
\bibitem{BTM} L. Baulieu and J. Thierry-Mieg, {\em Nucl. Phys.} {\bf 187}
(1982) 477~;
L. Baulieu, {\em Nucl. Phys.} {\bf B241} (1984) 557.
\bibitem{TM} J. Thierry-Mieg, {\em Phys. Lett.} {\bf 147B} (1984) 430.
\eject
\bibitem{BBH1} G. Barnich, F. Brandt and M. Henneaux, {\em Local
BRST cohomology of
Einstein gravity}, in preparation.
\bibitem{Ander} I. M. Anderson {\em The variational bicomplex}, Academic
Press (Boston : 1994)~; {\em Contemp. Math.} {\bf 132} (1992) 51.
\bibitem{BBH2} G. Barnich, F. Brandt and M. Henneaux, {\em
Variational bicomplex, characteristic cohomology and local BRST
cohomology}, in preparation.

\end{thebibliography}
\end{document}